\documentclass[12pt,preprint,a4]{iopart}
\usepackage[pdftex]{graphicx}
\usepackage[utf8]{inputenc}
\usepackage[english]{babel}
\usepackage{graphicx}
\usepackage{subcaption}

\usepackage[style=phys,biblabel=brackets,backend=biber,citestyle=numeric-comp,sorting=none]{biblatex}
\begin{document}

\title[Electrochemsitry]{Nanosecond and microsecond-pulsed plasma-in-liquid treated copper oxide surfaces}

\author{P. Pottkämper, N. Unteregge, S. Weller and A. von Keudell}

\address{Experimental Physics II - Reactive Plasmas, Ruhr-University Bochum, D-44780 Bochum, Germany}

\ead{Pia-Victoria.Pottkaemper@rub.de}

\begin{abstract}
Nanosecond and microsecond plasma-in liquid systems are explored to oxidize or regenerate a copper oxide surface in situ to serve as a catalyst for electrochemical CO$_2$ conversion. The plasma excitation generates H$_2$O$_2$ in the liquid, which induces the dissolution of Cu into Cu(OH)$_2$ and the recrystallization into Cu$_2$O nanocubes at the interface. The plasma performance of the two excitation schemes is analyzed, showing that the H$_2$O$_2$ production of nanosecond plasma is more efficient than of microsecond plasmas. The nature of the Cu$_2$O nanocubes is evaluated using electron microscopy and electrochemical characterization. 
\end{abstract}

\date{\today}
\maketitle
\section{Introduction}

Copper oxide surfaces are valuable catalysts in the electrolysis of CO$_2$ to hydrocarbons \cite{mistry_nanostructured_2016}. The conversion depends on the oxidation state and on the structure of the copper oxide surface \cite{zaza_welldefined_2022, reske_particle_2014}. The electrolysis process itself, however, causes performance deterioration \cite{grosse_dynamic_2018}. Therefore, an in-operando method is evaluated to re-oxidize the catalytic copper surface using nanosecond and microsecond plasmas, which could be operated in alternation between the electrolysis and oxidation during plasma treatment.

The oxidation of copper in an alkaline solution such as KOH follows several reaction steps, where OH$^-$ ions from the electrolyte react with copper and change its oxidation state, either to Cu$_2$O or Cu(I) oxide or to CuO or Cu(II) oxide. At first, an intermediate Cu(OH) or Cu(OH)$_2$ state is formed at the surface after electron transfer. These groups then decompose into CuO or Cu$_2$O and split off water molecules\cite{giri_electrochemical_2016,dechialvo_mechanism_1984}. This reaction is triggered during electrochemical growth of copper oxide by anodic oxidation providing reactive species from a plasma source. Alternatively the same reaction sequence could be initiated by a plasma source. The following reactions summarize the electron transfer at the interface:

\begin{eqnarray}
{\rm 2 Cu + 2 OH^-} &\longrightarrow& {\rm 2 Cu (OH) + 2 e^- } \\
{\rm 2 CuOH} &\longrightarrow& {\rm Cu_2O + H_2O}\\
{\rm Cu + 2 OH^-} &\longrightarrow& {\rm Cu(OH)_2 + 2 e^- \hspace{2cm} }\\
{\rm Cu(OH)_2} &\longrightarrow& {\rm CuO + H_2O}\\
{\rm Cu_2O + 6 OH^- + H_2O} &\longrightarrow& {\rm 2 Cu(OH)_4^{2-} + 2 e^-}\\
{\rm Cu + 4 OH^-} &\longrightarrow& {\rm Cu(OH)_4^{2-} + 2 e^-}
\end{eqnarray}


In this work, reactive species are generated through two different mechanisms: pulsed in-liquid plasma ignition on both a nanosecond and microsecond timescale. In both cases, a driven electrode and a grounded electrode are submerged inside the liquid and a high voltage is applied. Although the experimental setups of both plasmas are quite similar, their plasma mechanisms differ significantly: (i) On a microsecond scale, the liquid in the vicinity of the driven electrode tip evaporates due to Joule heating and a plasma is ignited in the gaseous phase inside the vapor-filled bubble. A streamer-like ignition then propagates through the liquid \cite{an_underwater_2007}; (ii) On the nanosecond scale, the plasma ignites directly inside the liquid, and different ignition scenarios have been postulated. Ignition may be triggered by electrostriction, where a high voltage with a short rise time creates nanovoids in the bulk liquid through electrostrictive stress. Inside such nanovoids the plasma can then ignite \cite{shneider_theoretical_2012} \cite{li_improved_2020}. Ignition may also occur through field effects \cite{grosse_ignition_2021} since the high electric field at the very sharp tip allows for direct field emission or field ionization depending on the polarity \cite{gomer_field_1972}. 

Regardless of pulse length and ignition mechanism, a variety of reactive species are generated once a plasma is formed. The plasma triggers water splitting reactions and ionization of its products, forming molecular species and ions such as H$_2$, O$_2$, H$^+$ and solvated electrons\cite{shih_chemical_2010}. These species recombine to more stable molecules and radicals such as H$_2$O$_2$. Eventually, a plasma-activated liquid (PAL) is generated. The PAL contributes to the oxidation and reduction of the copper surface, as described above. Based on the different ignition processes comparing microsecond and nanosecond plasmas in liquids, it is expected that the yield of reactive species and their effect on copper surfaces may also be different. The oxidation and reduction process of a copper surface exhibits a structural component since the growth of copper oxide crystals is induced by PAL treatment. We have previously described the growth of such crystals from PAL treatment\cite{pottkamper_plasma_2024} in a three-step process, where the copper is first oxidized into Cu(OH)$_2$ by hydrogen peroxide, dissolved into the liquid and then redeposits onto the surface where it is reduced to crystalline Cu$_2$O similar to the oxidation process of copper during chemical mechanical planarization (CMP) \cite{denardis_characterization_2006, denardis_studying_2010}. Since plasma ignition and, therefore, the properties of its plasma-activated liquid differ depending on applied pulse length, it is necessary to examine whether crystal growth is possible and whether the oxidation mechanisms differ for plasmas operated at microseconds or nanosecond high voltage pulses.

\section{Experimental}

\subsection{Plasma in liquid setups}

\begin{figure}[h]
    \centering
    \includegraphics[width=0.9\textwidth]{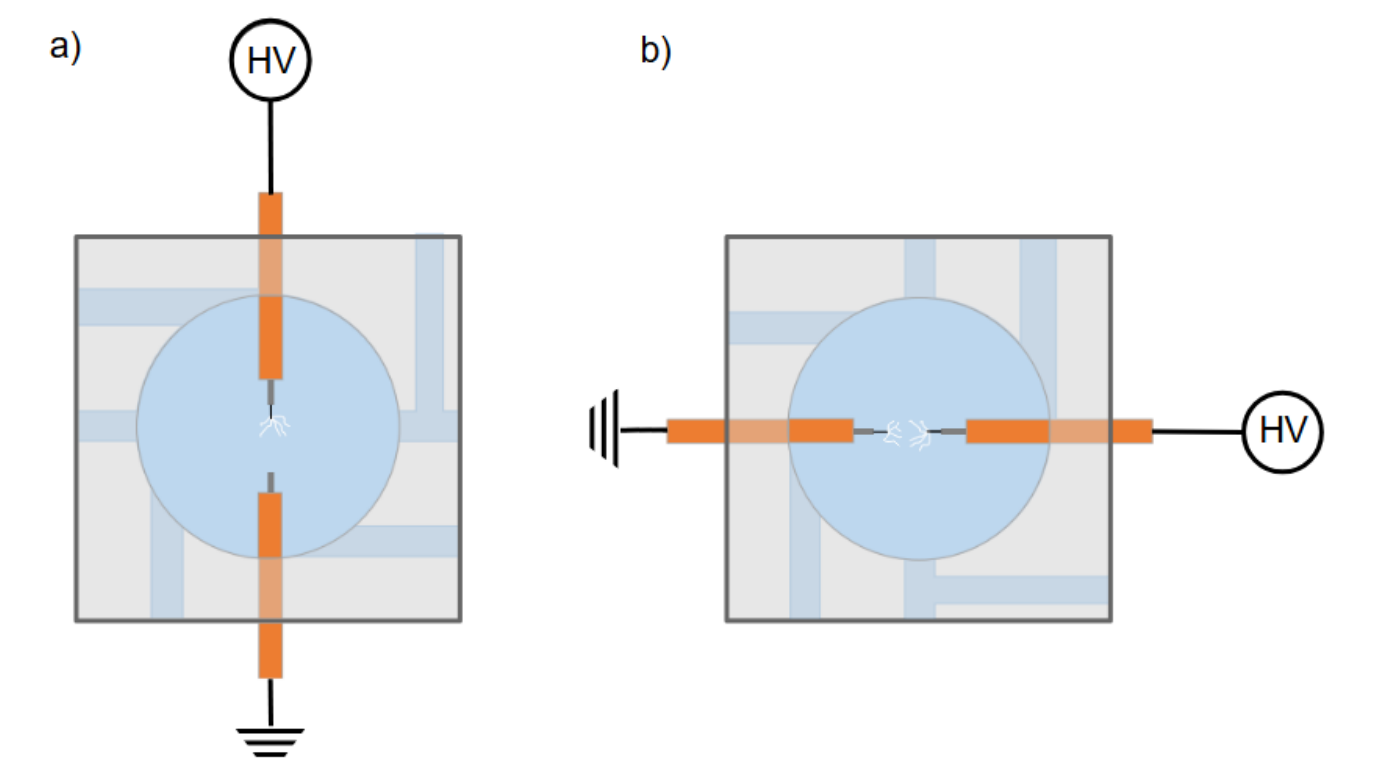}
    \caption{Comparison between (a) the nanosecond and (b) the microsecond pulsed in-liquid plasma reactor setups. The microsecond reactor contains two tungsten wire electrodes, both on the driven and grounded side in an effort to maximize the production of the reactive species.}
    \label{fig:Setups}
\end{figure}

Nanosecond and microsecond pulsed plasmas in liquids are used to prepare plasma-activated liquids (PAL) in separate reactors that are identical in construction and are made from PMMA and hold a liquid volume of 25 ml (see Figure \ref{fig:Setups}). The nanosecond setup is powered by a pulse generator producing a voltage pulse with a width of 12\,ns and a rising time of 2-3\,ns. The microsecond setup is powered by a pulse generator producing pulses with a width of 50\,$\mu$s and a rising time of 0.5\,$\mu$s. The plasma chambers are placed inside separate Faraday cages. In the nanosecond setup, the driven electrode consists of a 50\,${\rm\mu}$m diameter tungsten wire and the grounded electrode is made from stainless steel. In the microsecond setup, both the driven and the grounded electrodes are made from tungsten wires. As a liquid medium, distilled water with an electrical conductivity of 1\,${\rm\mu}$S cm$^{-1}$ and a pH of approximately 5.5 is used. 

\begin{figure}[h]
    \centering
    \includegraphics[width=0.5\textwidth]{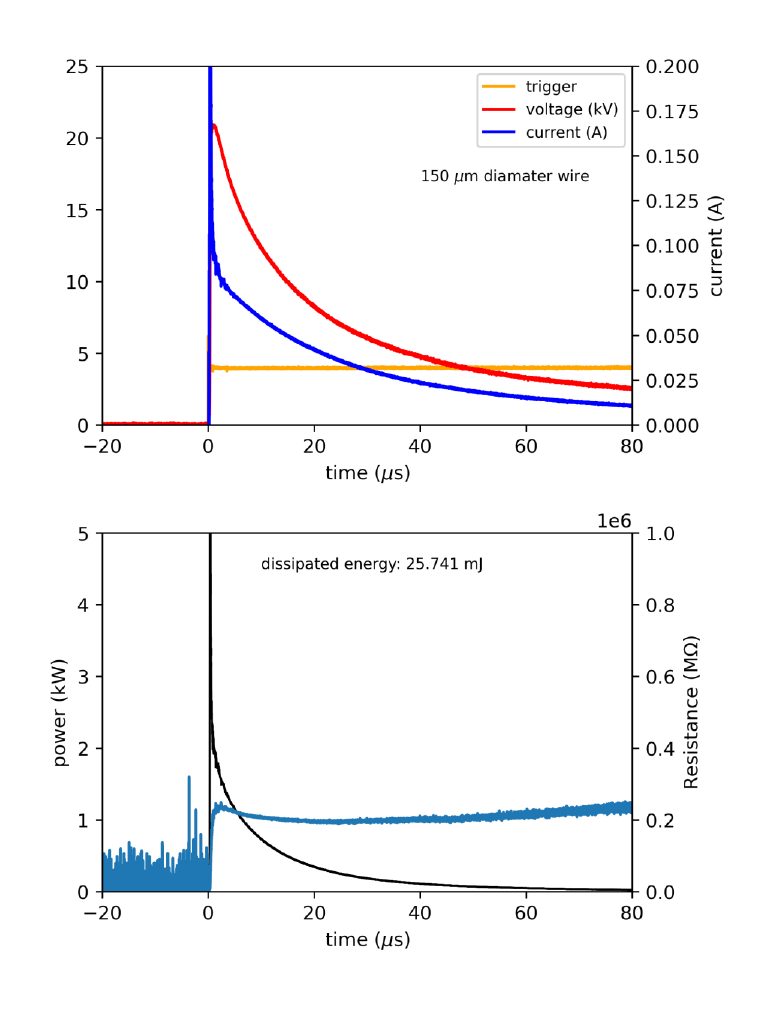}
    \caption{(a) VI characteristics of a microsecond plasma pulse using a voltage of 21 kV and a wire with a diameter of 150 $\mu$m; (b) power (black line) and Ohmic resistance of the system (blue line).}
    \label{fig:vi}
\end{figure}

Fig. \ref{fig:vi}(a) shows a typical voltage-current curve for the $\mu$s plasma. From this, the absorbed power and the resistance of the discharge is deduced with $R=U/I$ and $P=U \cdot I$, as shown in Fig. \ref{fig:vi}(b). It is striking to see that the current and voltage follow in phase and that the Ohmic resistance stays almost constant over time at 0.2 M$\Omega$. If we integrate the instantaneous power over time, we obtain a dissipated energy of 25.7 mJ. The resistance of 0.2 M$\Omega$ can be connected to the measured conductivity of $\sigma$ = 1 $\mu$S/cm by estimating the volume the current has to pass from the powered electrode to the grounded counter electrode. The resistance is given as $R = \frac{L}{\sigma r^2 \pi}$ with $L$, the length of a cylinder with radius $r$. If we set $L$ the length between the two electrodes, we obtain a radius $r$ = 1.2 cm.

\begin{figure}[h]
    \centering
    \includegraphics[width=0.7\textwidth]{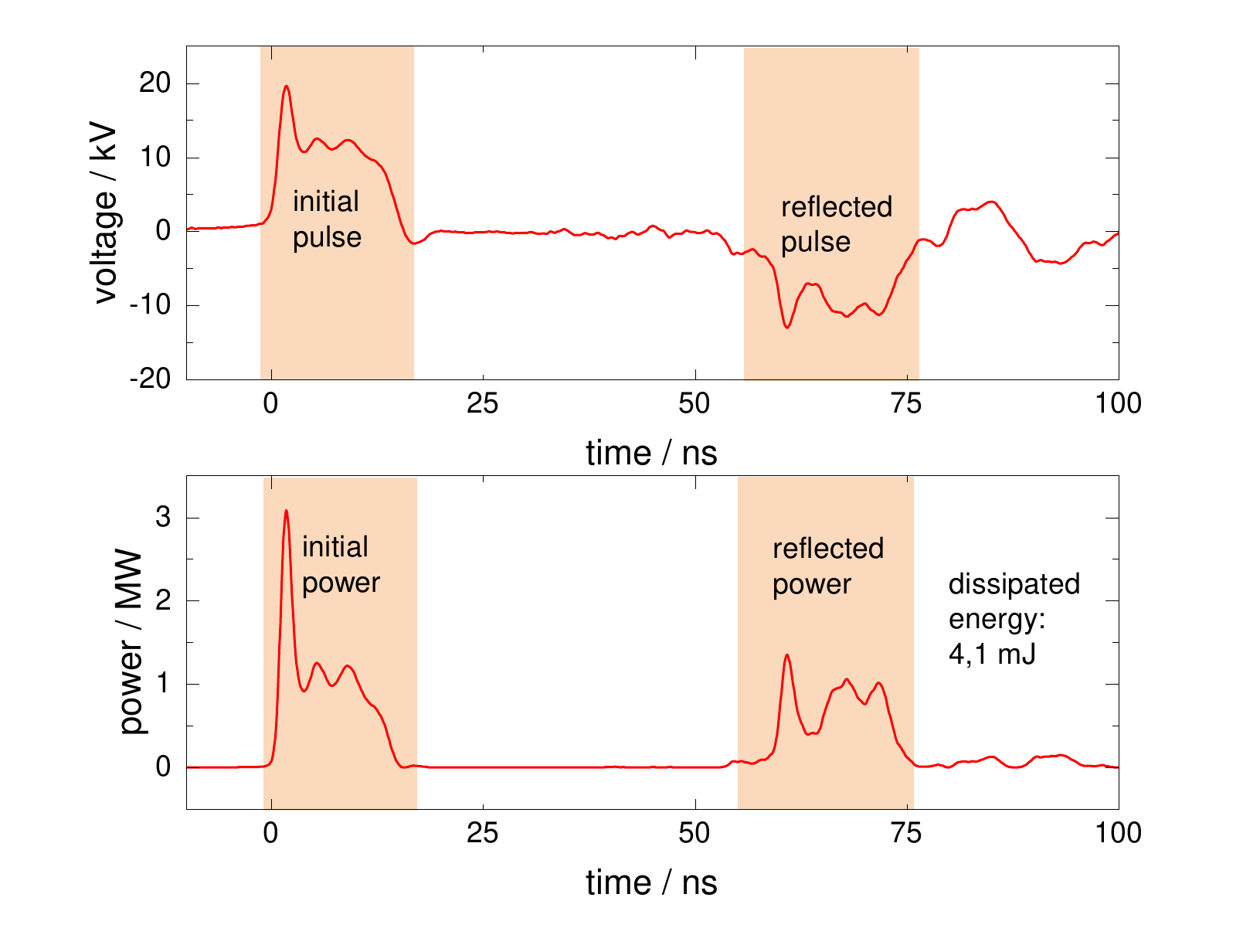}
    \caption{characteristic voltage energy and power curve of the ns pulsed in-liquid plasma ignition.}
    \label{fig:vi2}
\end{figure}

Fig. \ref{fig:vi2} shows a voltage and power curve of the ns plasma. The voltage is indirectly measured using a back current shunt (BCS) to record an incident and a reflected pulse. Due to the mismatch of the impedance of the setup, the pulse is partially reflected and travels back and forth between the electrode tip and the power generator. By subtracting the backwards from the forward travelling high voltage pulse, the dissipated energy can be deduced. For this, the incident and reflected power are integrated over time to yield the incident and reflected energies. By subtracting the reflected energy from the incident energy, we yield the dissipated energy in the order of 4.1 mJ at a voltage of 20 kV. These dissipated energies are one order of magnitude smaller than for the $\mu$s plasma. 

\subsection{Sample preparation}

Copper samples were prepared using High Power Impulse Magnetron Sputtering (HiPIMS) and a copper target. A silicon oxide wafer is used as a substrate placed opposite to the copper target at a distance of 7 cm. The coating process uses argon as sputter gas at a pressure of 0.5\,Pa. The HiPIMS power supply is operated at 740\,V, 110\,A, 1\,kW, 35\,Hz and a pulse time of 50 µs, yielding a deposition rate of 1.3 nm/s on the substrate. Thereby, 100 nm copper layers are deposited within 130 s. After preparation, the samples are removed from the system and are inadvertently exposed to ambient air, creating an initial oxide layer. The samples are then treated with the PAL by placing a drop of about 1 ml of PAL on top for approximately 4 hours.

\subsection{Analysis of plasma-treated liquids}

As initial liquid samples, distilled water with a conductivity of 2$\mu$S/cm and a 0.05 mM KOH solution with a conductivity of 6$\mu$S/cm were investigated. The PAL samples were prepared for one hour by plasma treatment at +20 kV, 10 Hz using both the microsecond and nanosecond-pulsed plasmas. Directly after PAL generation, any temperature increase during preparation and any change in electrical conductivity are measured using a combined thermocouple and conductivity meter (Greisinger GLF100). Then, the PAL was removed from the plasma reactor and a droplet was placed on the sample and left there for approximately four hours. 

The H$_2$O$_2$ content of the PAL was investigated by photometric spectroscopy using the Merck 118789 assay kit (Spectroquant) sensitive in the range of 0.015 to 6.00 mg/l H$_2$O$_2$. The PAL is diluted with distilled water in ratio 1/9 and then thoroughly mixed with the two phases of the assay. The samples are analyzed in a photometer adjusted for the absorption wavelength corresponding to the assay. The concentrations are calculated using a calibration curve determined from solutions with known H$_2$O$_2$ concentrations. In addition, PAL solutions with known concentrations were analyzed to verify the accuracy of the calibration.

The quantification of the H$_2$O$_2$ concentration is difficult because H$_2$O$_2$ is not a stable species and can react back to water over time, especially at elevated temperatures. Therefore, the quantification of plasma-generated H$_2$O$_2$ has been measured as quickly as possible. From a time series, a decay constant of 0.0015 mM/h has been found.

\subsection{Analysis of plasma treated surfaces samples}

\begin{figure}[h]
    \centering
 \includegraphics[width=0.6\textwidth]{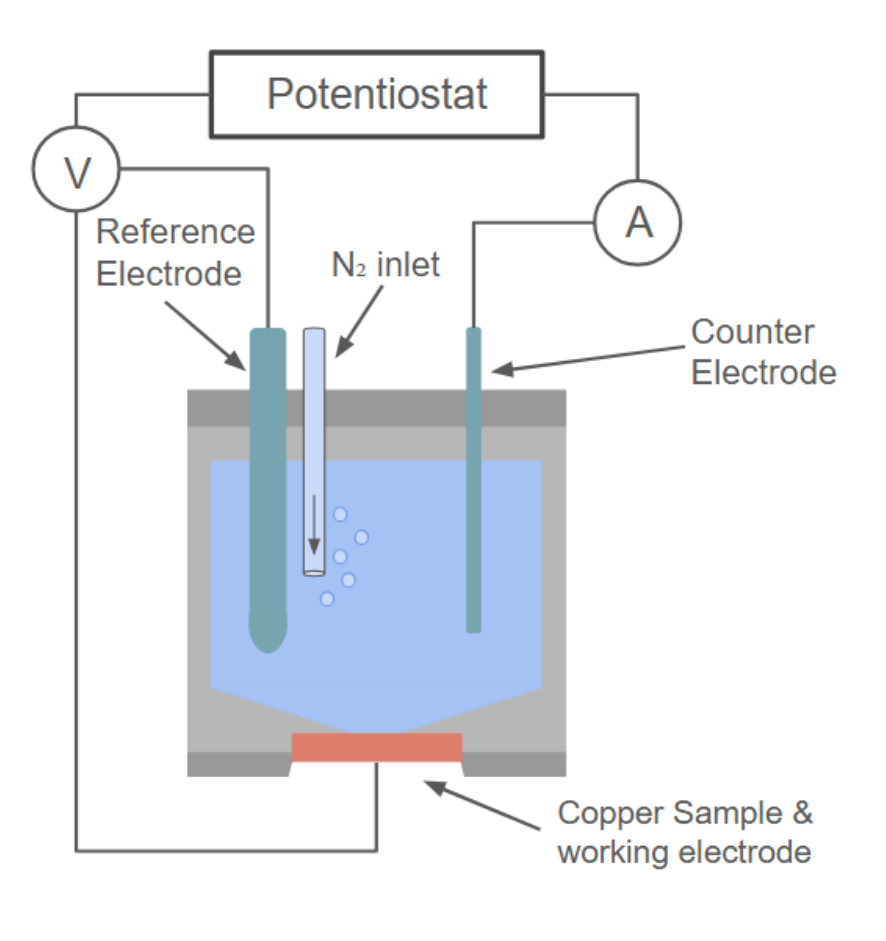}
 \caption{Sketch of the cyclic voltammetry setup. The copper sample is placed at the bottom of the reactor and acts as the working electrode. Both the reference and counter electrode are placed in the electrolyte. An inlet for gaseous nitrogen is located at the top of the reactor to degas the electrolyte before the measurement. }
    \label{fig:exp_setup_CV}
\end{figure}

The influence of PAL on the samples was investigated using cyclic voltammetry (CV) and scanning electron microscopy (SEM). Since the CV cycles dissolved the copper oxide surface layers induced by the PAL, separate samples were prepared for CV and SEM measurements. A scheme of the CV cell is shown in Fig. \ref{fig:exp_setup_CV}. At first, the treated samples are installed as the working electrode at the bottom of the CV cell. The clean cell is then filled with an electrolyte. For this purpose, a solution consisting of 0.01 M KOH and 0.1 M K$_2$SO$_4$ was used. To prevent a background signal from the oxygen in the electrolyte during the measurement, the electrolyte was degassed using a weak nitrogen flow for 20 minutes. The N$_2$ inlet was removed before the measurement was performed. Then, the open-circuit voltage (OCV), a cyclic voltammogram (CV) and an electrical impedance measurement (EIS) were recorded. The EIS data was used to correct the Ohmic drop in the voltammogram. The Ohmic drop occurs due to the inherent resistance of the electrolyte, which causes a shift in the measured curve and needs to be accounted for.

The CV curve is characteristic of surface species being reduced (or oxidized) at a certain voltage E$_{we}$. One may distinguish amorphous, intermediate and crystalline Cu$_2$O \cite{gil_electrochemical_2009} surface groups. CuO or Cu(OH)$_2$ surface species show similar reduction potentials \cite{dong_cyclic_1992} and can not be distinguished in CV measurements. According to \cite{gil_electrochemical_2009} these characteristic peaks are at -0.5 V (E$_{we}$) at the working electrode for the reduction process of CuO to Cu$_2$O, at around -0,65 V E$_{we}$ for the reduction of amorphous Cu$_2$O to Cu, at -0.8 V V E$_{we}$ a broad peak can be seen for the reduction of intermediate Cu$_2$O to Cu and finally at -0.1 V E$_{we}$ crystalline Cu$_2$O reduces to pure Cu.

\newpage

\section{Results and discussion}

\subsection{Plasma performance}

At first, we regard the difference in plasma performance of the ns-plasmas and $\mu$s plasmas. Fig. \ref{fig:dissEnergy} shows the dissipated energies as they change over time of plasma operation treating distilled water using the microsecond and nanosecond pulsed power supplies. When using the $\mu$s-plasma, the dissipated energy is initially at 16 mJ, which is much larger than the 3 mJ delivered by the nanosecond-pulsed plasma. For longer operation times, the dissipated energy for ns-plasma remains unaltered, whereas the dissipated energy steadily increases to 32 mJ after 70 minutes of $\mu$s plasma operation time. 

At longer treatment times, the $\mu$s plasma abruptly ignites also at the grounded electrode. The dissipated energies increase abruptly to 92 mJ after 80 minutes and continues to increase from there. This transition corresponds to the transition from streamer mode to a spark mode, where the plasma transitions into an arc in between the electrodes, which causes an intense plasma emission and much higher currents.  This transition may be triggered at a specific temperature and conductivity, where the bubble formation by plasma ignition amplifies itself in subsequent plasma pulses. This transition occurs spontaneously depending on the distance between the two electrodes and the random accumulation of gas bubbles. Therefore, the moment in time when this transition occurs varies between experiments. In subsequent measurements it has been found, that the hydrogen peroxide production efficiency is much higher in the initial streamer-like discharge mode. With the transition into the spark mode the hydrogen peroxide concentration decreases abruptly, presumably due to the thermal destruction of H$_2$O$_2$. Therefore in the following text emphasis will be placed on the streamer-like discharge mode.

The significant difference in the initial energies are attributed to the very different ignition mechanisms in both plasmas: (i) in the case of the ns-plasma the dissipated energies are very small so that the system remains almost unchanged and the plasma operation is repetitive; (ii) in case of the $\mu$s plasma, the higher dissipated energies lead to the accumulation of bubbles during plasma operation, as can be observed in the experimental setup itself (not shown). Consequently, at later times, ignition does not only occur in the newly created bubbles at the very tip of the electrode but also inside bubbles which have been created in previous $\mu$s pulses. Thereby, the affected plasma volume grows over time and thus the dissipated energy. These different trends can also be seen when considering the change in liquid temperature and conductivity:
 
\begin{figure}[h]
    \centering
    \includegraphics[width=0.7\textwidth]{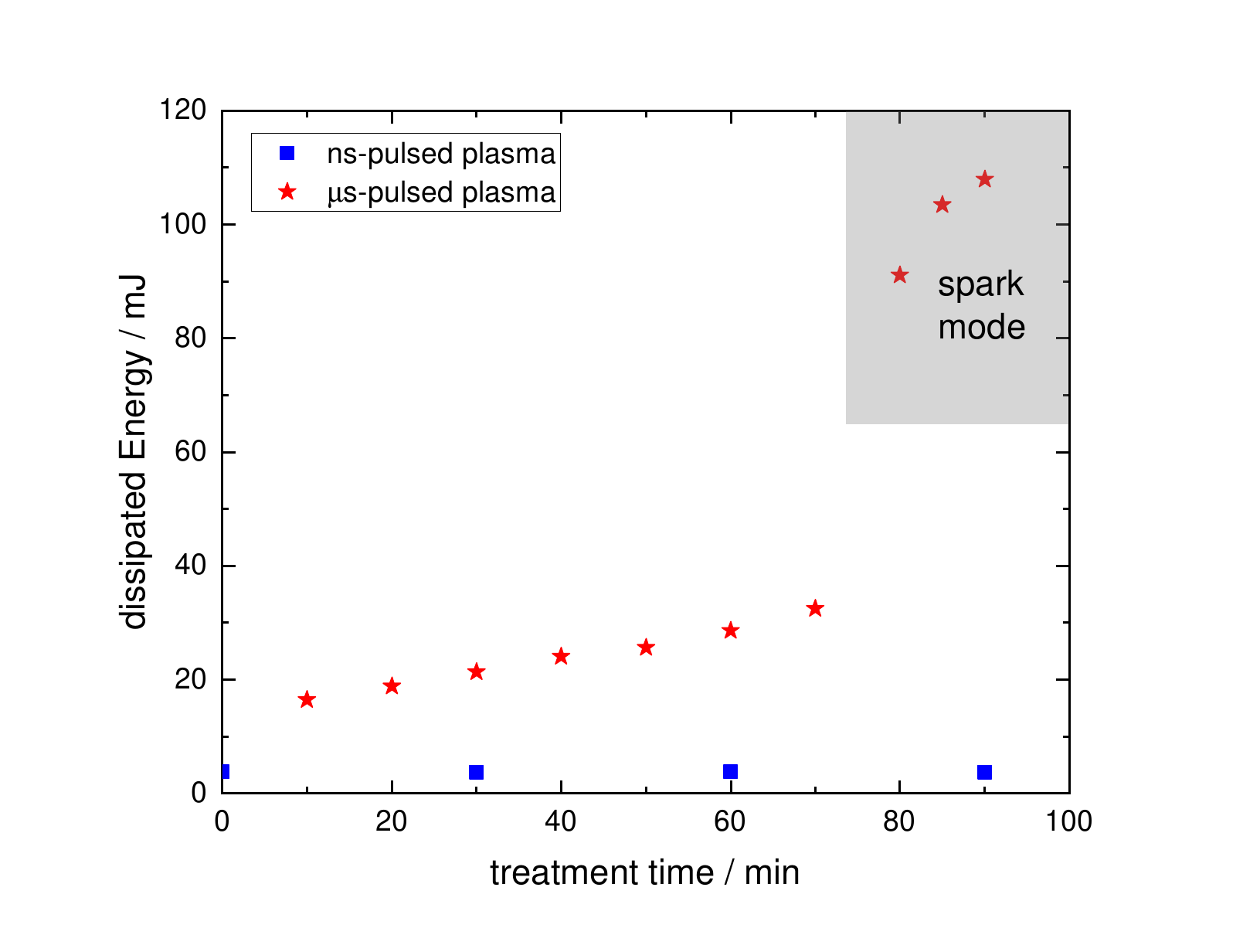}
    \caption{Dissipated energies of the microsecond- and nanosecond pulsed plasma setups at 20 kV over time.}
    \label{fig:dissEnergy}
\end{figure} 

\begin{figure}[h]
    \centering
    \includegraphics[width=0.7\textwidth]{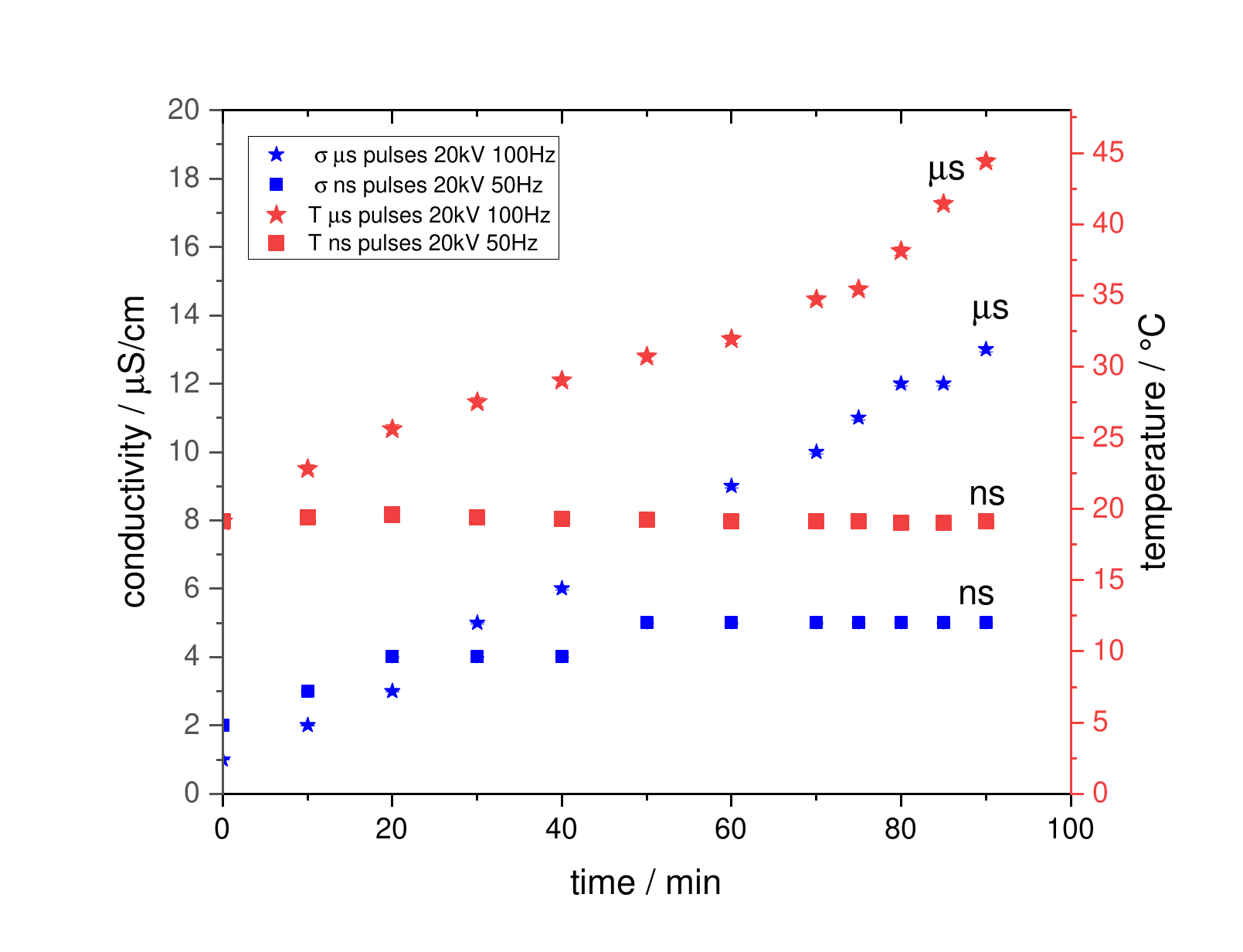}
    \caption{Change in conductivity in the micro and nanosecond pulsed treated liquid (blue points) during an experiment using a voltage of 20 kV for 90 min at 100 Hz and 50 Hz, respectively. Change in the temperature of the liquids (red points) at the same settings.}
    \label{fig:cond}
\end{figure}

Fig. \ref{fig:cond} shows the change in temperature of distilled water exposed to a microsecond plasma at a voltage of 20 kV at 100 Hz and a nanosecond plasma at 20 kV and 50 Hz, both for 90 minutes. The temperature increases from 20 °C to 45 °C in the $\mu$s plasma case and remains at 19 °C for the ns plasma case. The upper limit for the change in temperature of the liquid treated by the $\mu$s plasma is derived from standard thermodynamics, assuming a perfectly insulated system. With the heat capacity of liquid water ($C_p$ = 4189 J/kg/K), the volume of water (30 ml, corresponding to 0.03 kg), we obtain a temperature increase of 71 K for an accumulated dissipated energy of 9000 J over 1 hour. This is only an upper limit because thermal conduction to the ambient will largely reduce this temperature increase. 

Fig. \ref{fig:cond} also shows the change in conductivity of distilled water for both plasma treatments.  In the $\mu$s case, the conductivity increases from 1 $\mu$ S/cm to 14 $\mu$ S/cm, while in the ns case, it only increases slightly from 2 $\mu$ S/cm to 5 $\mu$ S/cm. The large dissipated energy in the case of the $\mu$s plasma causes the generation of more ions, leading to a higher conductivity at the end. This change in conductivity can be roughly assessed as follows. Suppose we assume that the dissipated energy is converted into generating OH$^-$ ions in the liquid. In that case, the dissipated energy of 1 mJ converts into 1.25 $\cdot$ 10$^{15}$ ions if we use the H-OH dissociation energy 5.11 eV as the formation energy of an OH$^-$ ion in water. This leads to a change in concentration in the 30 ml liquid compartment of $n_{OH^-}$ = 4.1 $\cdot$ 10$^{13}$ m$^{-3}$. When using the mobility of OH$^-$ ions of $\mu_e$ = 20.8 $\cdot$ 10$^{-8}$ m$^2$ S$^{-1}$V$^{-1}$, we obtain a change in conductivity $\sigma = \mu_{OH}n_{OH}e$ of 0.3 $\mu$S/cm per hour of plasma operation at 100 Hz per 1 mJ dissipated energy per pulse. Therefore, at a dissipated power of 30 mJ, we expect an increase in conductivity of 30 $\mu$S/cm, which is in rough quantitative agreement to the increase by 13 $\mu$S/cm for the $\mu$s plasma.

\subsection{Production efficiency of H$_2$O$_2$}

\begin{figure}[h]
    \centering
    \includegraphics[width=0.7\textwidth]{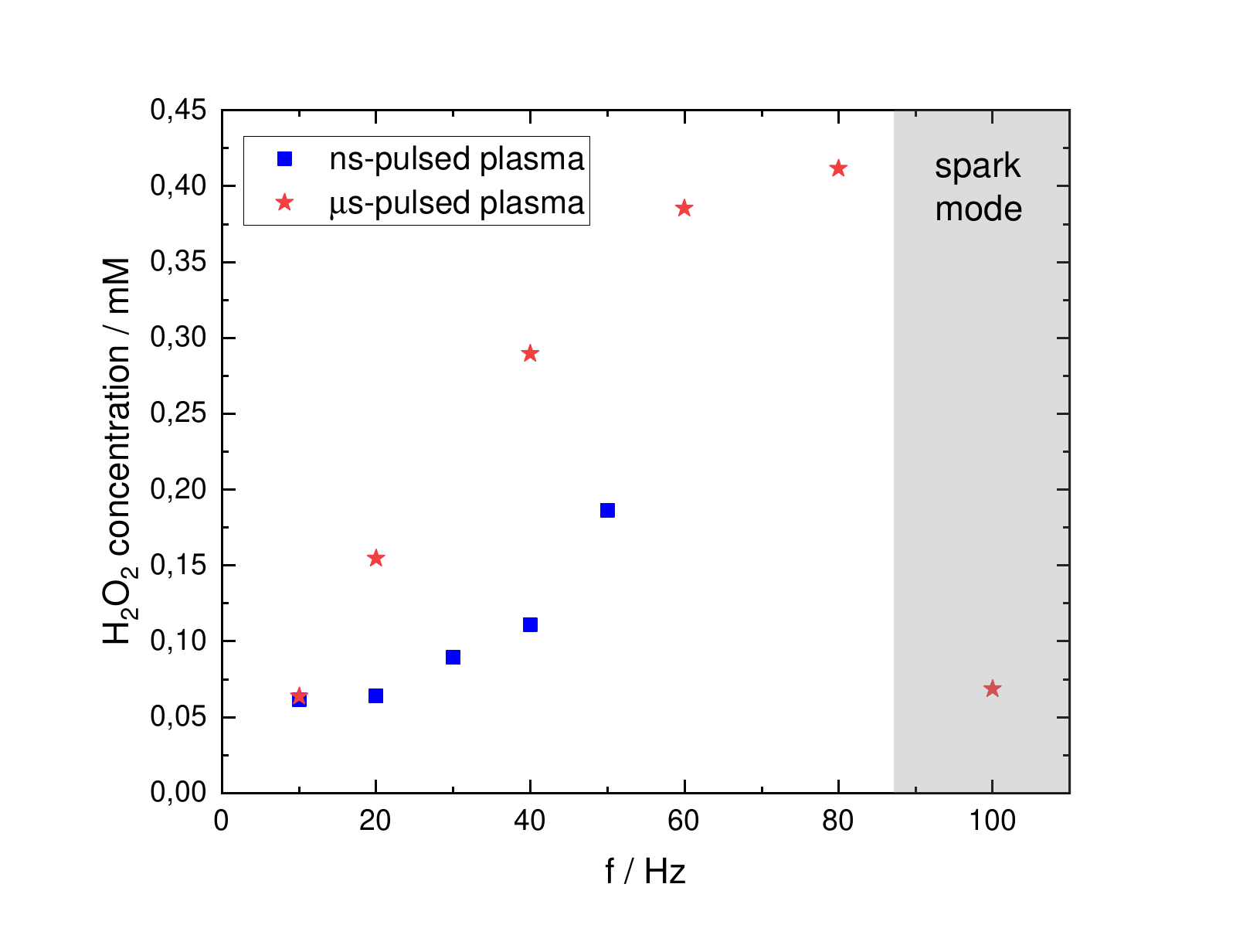}
    \caption{Hydrogen peroxide concentrations from the PAW produced with microsecond- and nanosecond-pulsed plasma at 20kV for one hour at varying frequncies.}
    \label{fig:h2o2concentration}
\end{figure}

In Figure \ref{fig:h2o2concentration} the frequency dependent hydrogen peroxide concentrations produced with the microsecond- and nanosecond-pulsed plasmas in distilled water are compared. The water was treated for one hour at 20 kV in both measurements. The ns plasma produces about 0.06 mM H$_2$O$_2$ at 20 kV and 10 Hz at a treatment time of 1 hour in the 30 ml volume. The concentration then steadily rises to about 0.2 mM at 50 Hz. The H$_2$O$_2$ production of the nanosecond setup is quantified, yielding a production efficiency of 0.5 g/kWh. 

This production efficiency has already been investigated previously \cite{chauvet_chemistry_2020} with the production increasing linearly with time, voltage and frequency, yielding an efficiency of 2 g/kWh. This was slightly higher in the past, which might be attributed to the different measurement methods. This efficiency is similar to other plasma methods in the literature, as listed in Tab. \ref{tab:h2o2generation}. 

In the beginning, the $\mu$s plasma also produces about 0.06 mM H$_2$O$_2$ at 20 kV and 10 Hz at a treatment time of 1 hour in the 30 ml volume. The production rises more steeply to about 0.4 mM at 80 Hz. Initially it seems that more hydrogen peroxide can be generated with longer plasma pulses. However, with a dissipated energy of about 30 mJ per pulse at higher treatment times, we obtain an energy efficiency of about 0,2 g/kWh for the $\mu$s plasma. This is lower in comparison to the energy efficiency of the nanosecond-pulsed plasma. Additionally, once the microsecond system enters the spark mode, where the dissipated energies rise abruptly to around 100 mJ, the H$_2$O$_2$ production sharply decreases. This is explained by the increase in temperature. It has been established, that the lifetime and production efficiency of hydrogen peroxide decreases at higher temperatures \cite{stefanic_temperature_2002}. 
As described above, no significant increase in temperature in the nanosecond case can be seen where the overall increase in temperature in the microsecond case is significant and very likely leads to a thermal dissociation of the H$_2$O$_2$.
In the spark mode, the temperature of the liquid rises suddenly and much of the H$_2$O$_2$ is destroyed. The production then continues at a much lower rate.

Even in the conventional streamer-like mode at lower frequencies, $\mu$s-pulsed plasma has a lower energy efficiency than similar plasma setups listed in Tab. \ref{tab:h2o2generation}. This may be attributed to the very low absolute cw power in our experiments of a few W only, whereas other plasma experiments in Tab.  \ref{tab:h2o2generation} operate at a few 10 W cw power. 

In addition, Tab. \ref{tab:h2o2generation} also list H$_2$O$_2$ production from water vapor admixed to a helium-diluted plasma jet\cite{schuttler_tuning_2025} yielding an energy efficiency of only 0.2 mg/kWh which is a factor 250 smaller in comparison to the $\mu$s plasma. This can easily be explained by the fact that the factor 250 corresponds roughly to the helium dilution ratio. Therefore, one may argue that the plasma excitation in the RF jets and microsecond plasmas is rather similar but the plasma in a liquid system provides only water as a collision partner. The best performance is provided by the ns plasmas, with a yield which is again 2-10 times higher than the microsecond plasma. In the case of the nanosecond plasma, the excitation is very fast, followed by a strong cooling of the reactive species, which preserved H$_2$O$_2$ after its production. In contrast, it has been found that in high power discharges, where thermal effects are significant, the hydrogen peroxide concentrations decrease due to the thermal degradation of H$_2$O$_2$ \cite{locke_review_2011}. 

\begin{table}[]
    \centering
    \begin{tabular}{|p{6cm}|c|l|} \hline
         Plasma source & energy yield [mg/kWh] & reference \\ \hline\hline
         capillary jet above liquid & 0.2 & \cite{schuttler_tuning_2025} \\
         $\mu$s discharge in liquid, high power& 1000-2100 & \cite{shih_chemical_2010,yamabe_water_2005}  \\
         ns pulses in liquid & 2000 & \cite{chauvet_chemistry_2020} \\
         ns pulses in liquid & 500-1000 & this work \\
         $\mu$s pulses in liquid,low power & 200 & this work \\ \hline
    \end{tabular}
    \caption{Energy efficiencies of various methods for H$_2$O$_2$ generation at similar experimental configurations.}
    \label{tab:h2o2generation}
\end{table}

\subsection{Surface modification by plasmas}

\begin{figure}[h]
    \centering
    \includegraphics[width=0.6\textwidth]{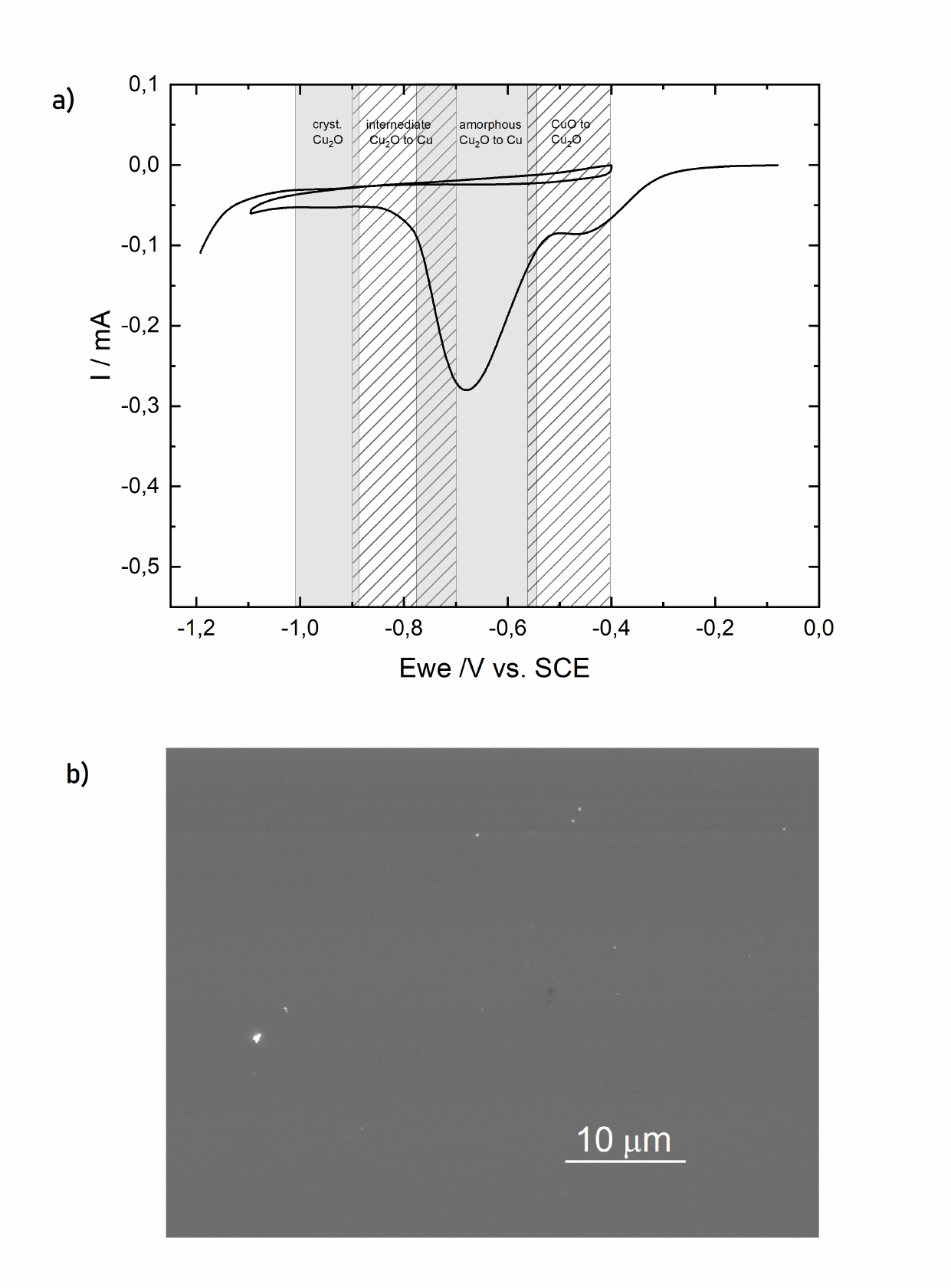}
    \caption{(a) CV curve of an untreated copper sample, (b) SEM images of an untreated copper sample}
    \label{fig:pureCu} 
\end{figure}

Fig. \ref{fig:pureCu}a shows a CV curve of an untreated copper sample with its natural oxide that grows from the ambient after preparation as a reference. The reduction reaction occurs at -0.65 V $E_{we}$, indicating that the copper oxide is in an amorphous state. No peak at lower voltages $E_{we}$ indicating crystal growth can be seen. 
A small signal originating from the reduction of either CuO or Cu(OH)$_2$ can also be seen at -0.45 V $E_{we}$. Since the two reduction potentials are similar, they can not be distinguished. However, previous experiments using this setup have shown that little to no CuO is present on the samples \cite{pottkamper_plasma_2024}. It is known that CuO can only be formed at high temperatures far beyond room temperature \cite{delossantosvalladares_crystallization_2012}. In addition, it can be seen that the reduction peaks are absent in the second CV cycle, indicating that the electrochemical reaction eventually removes the surface oxides at the top most layers. This is attributed to the fact that the depletion of the topmost surface in the first cycle is enough to render the surface oxide-free in the second cycle. 

Fig. \ref{fig:pureCu}b shows an SEM image of the untreated sample, where no oxide crystals can be identified. This corresponds well to the amorphous Cu$_2$O peak in the CV curve.

\begin{figure}[h]
    \centering
    \includegraphics[width=0.6\textwidth]{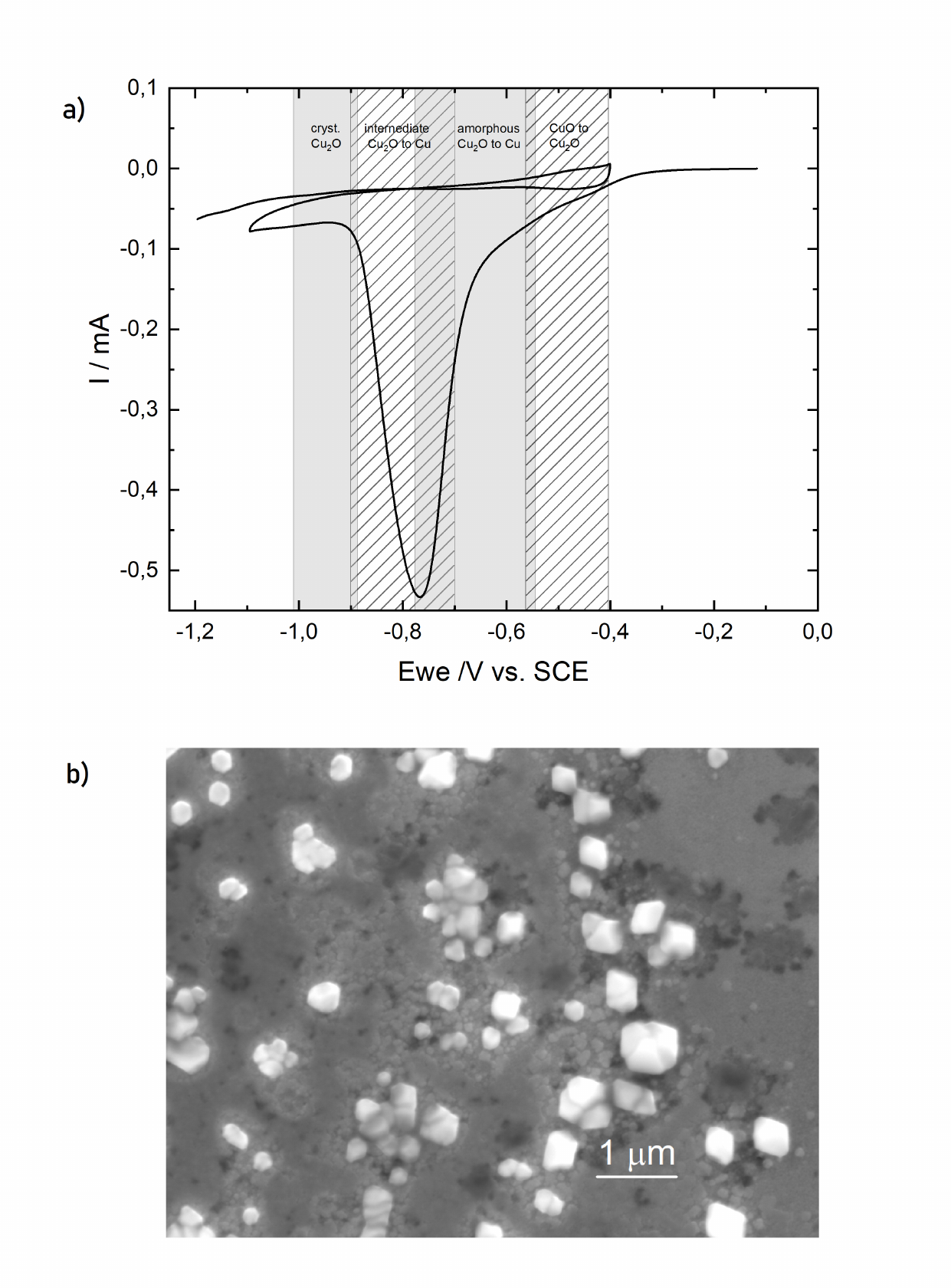}
    \caption{(a) Voltammogram of a copper sample treated with distilled water, (b) SEM image of a copper sample treated with distilled water.}
    \label{fig:distwater}    
\end{figure}

Fig. \ref{fig:distwater}a shows a CV curve of a copper sample treated with distilled water. The reduction peak is now at -0.8 V $E_{we}$, indicating an intermediate state of copper oxide. The shoulder at -0.4 V $E_{we}$ disappeared, indicating the absence of any CuO groups. Still, no crystalline signal at -1 V $E_{we}$ is visible in the CV curves. This is in contrast to the corresponding SEM image \ref{fig:distwater}b where some crystal formation is observed. The brighter areas in the SEM images have a diameter of 1$\mu$m, but most are significantly smaller. These areas also show no defined crystalline structure.

\begin{figure}[h]
    \centering
    \includegraphics[width=0.6\textwidth]{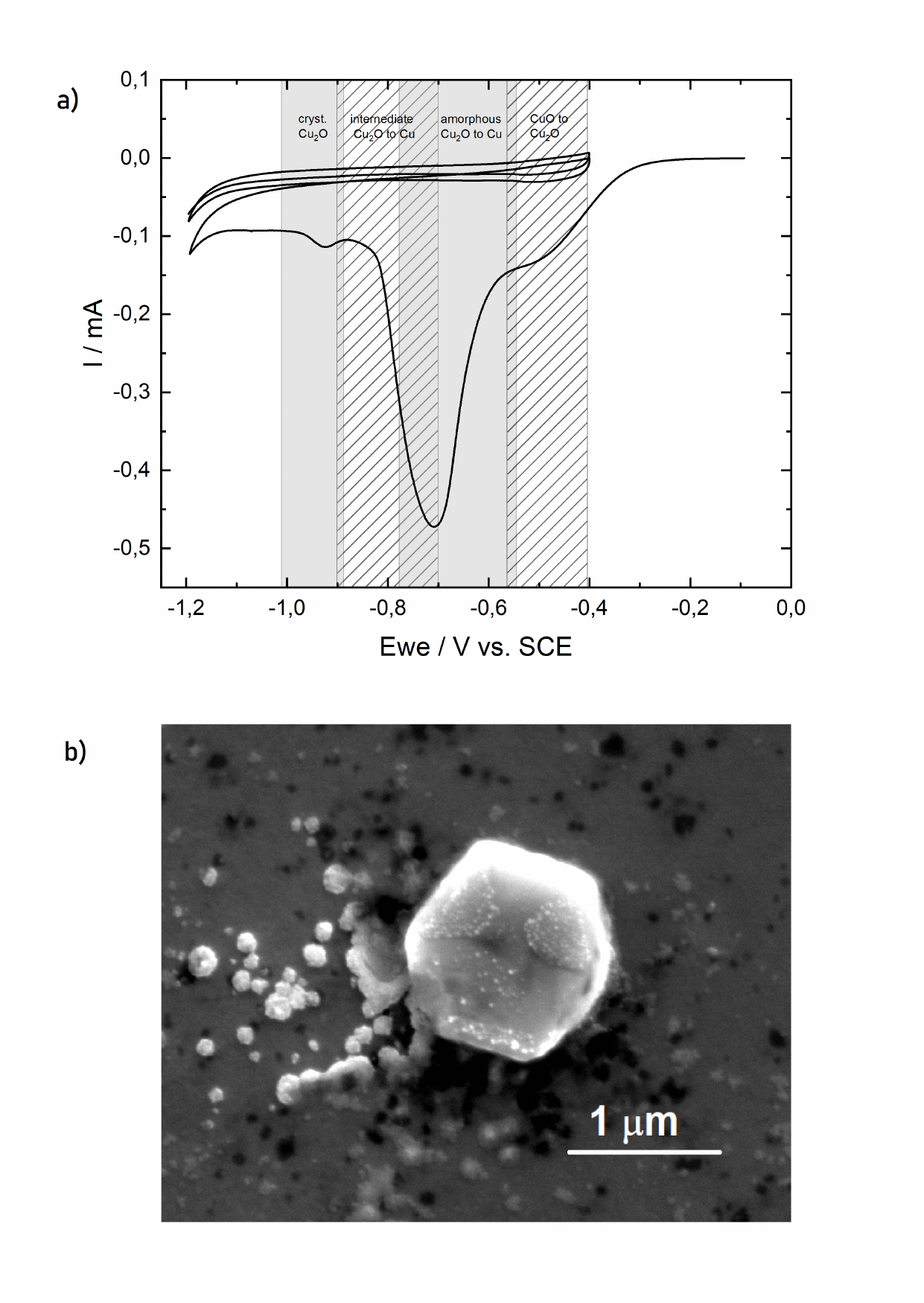}
    \caption{(a) Voltammogram of a copper sample treated with plasma-activated water, (b) SEM image of a copper sample treated with plasma-activated water.}
    \label{fig:PAW}
\end{figure}

Fig. \ref{fig:PAW}a shows the CV curve of a sample treated with PAL. The microsecond-pulsed plasma was used in this case. Here, the large peak at -0.8 V $E_{we}$ indicates that the copper oxide still seems to be in the intermediate state. However, a small signal in the crystalline range at at -1 V $E_{we}$ is visible. Additionally, there is a signal in the amorphous range at -0.6 V $E_{we}$, indicating the presence of H$_2$O$_2$ in the liquid which causes the formation of Cu(OH)$_2$. This leads to the conclusion that Cu$_2$O crystals have formed as confirmed by the SEM image in Fig. \ref{fig:PAW}b. 

Well-defined crystals can be found on those samples. Often times, a crystal can be found surrounded by a uniform area with no visible oxidation. This may explain the increase in the amorphous signal. From the SEM measurements, it has been found that the larger crystals seem to detach from the samples easily. Since the preparation for the CV measurements requires a distinct handling of the samples, any loss of surface crystals cannot be avoided.

Next, we will examine the electrochemical characterization of samples treated with plasma-activated liquids using microsecond plasmas. Despite its lower energy efficiency, the absolute H$_2$O$_2$ concentration in the microsecond-generated liquids is higher than in the ns-treated liquids.

\begin{figure}[h]
    \centering
    \includegraphics[width=0.6\textwidth]{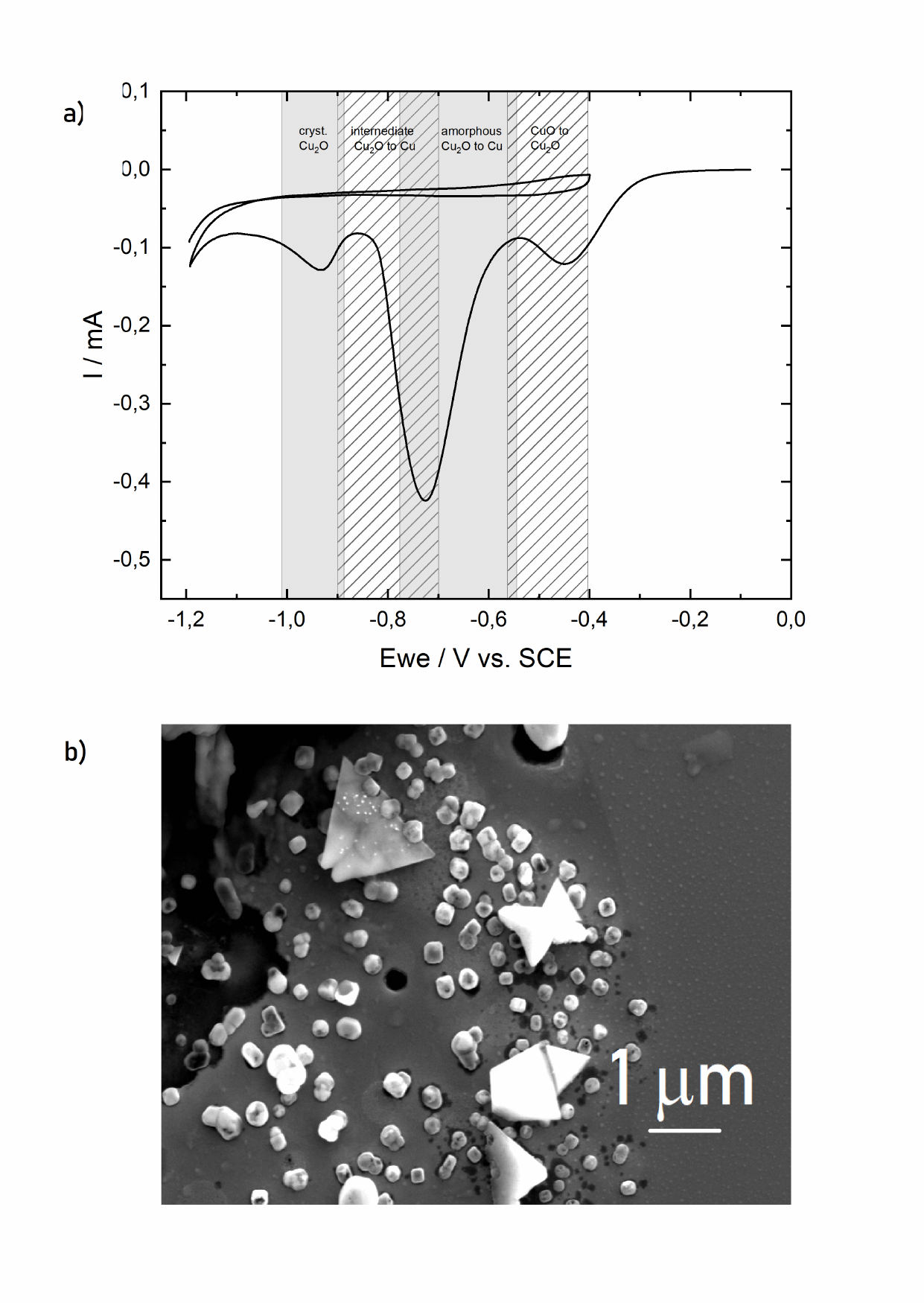}
    \caption{(a) Voltammogram of a copper sample treated with microsecond pulsed plasma-activated water,(b) SEM image of a copper sample treated with the same solution.}
    \label{fig:usPAW}
\end{figure}

Figure \ref{fig:usPAW}a shows a CV curve for a sample being treated with PAL distilled water as activated using the $\mu$s pulse generator. The sharp peak at -0.8 V $E_{we}$ indicates that the Cu$_2$O at the surface is the intermediate state but, also, a distinct peak at -0.5 V $E_{we}$ indicating the CuO and at -1.0 V $E_{we}$ indicating crystalline Cu$_2$O range can be seen. This is reasonable since the absolute H$_2$O$_2$ concentration in the $\mu$s treated liquids is higher than for the ns plasma-treated liquids. 

Fig. \ref{fig:usPAW}b shows the SEM image indicating a well-defined copper oxide crystals (similar to the ns case in Figure \ref{fig:PAW}b). These results indicate, that is is possible to elicit Cu$_x$O crystal formation through the treatment of the liquid using the $\mu$s-pulsed plasma as well. 

\begin{figure}[h]
    \centering
        \includegraphics[width=0.6\textwidth]{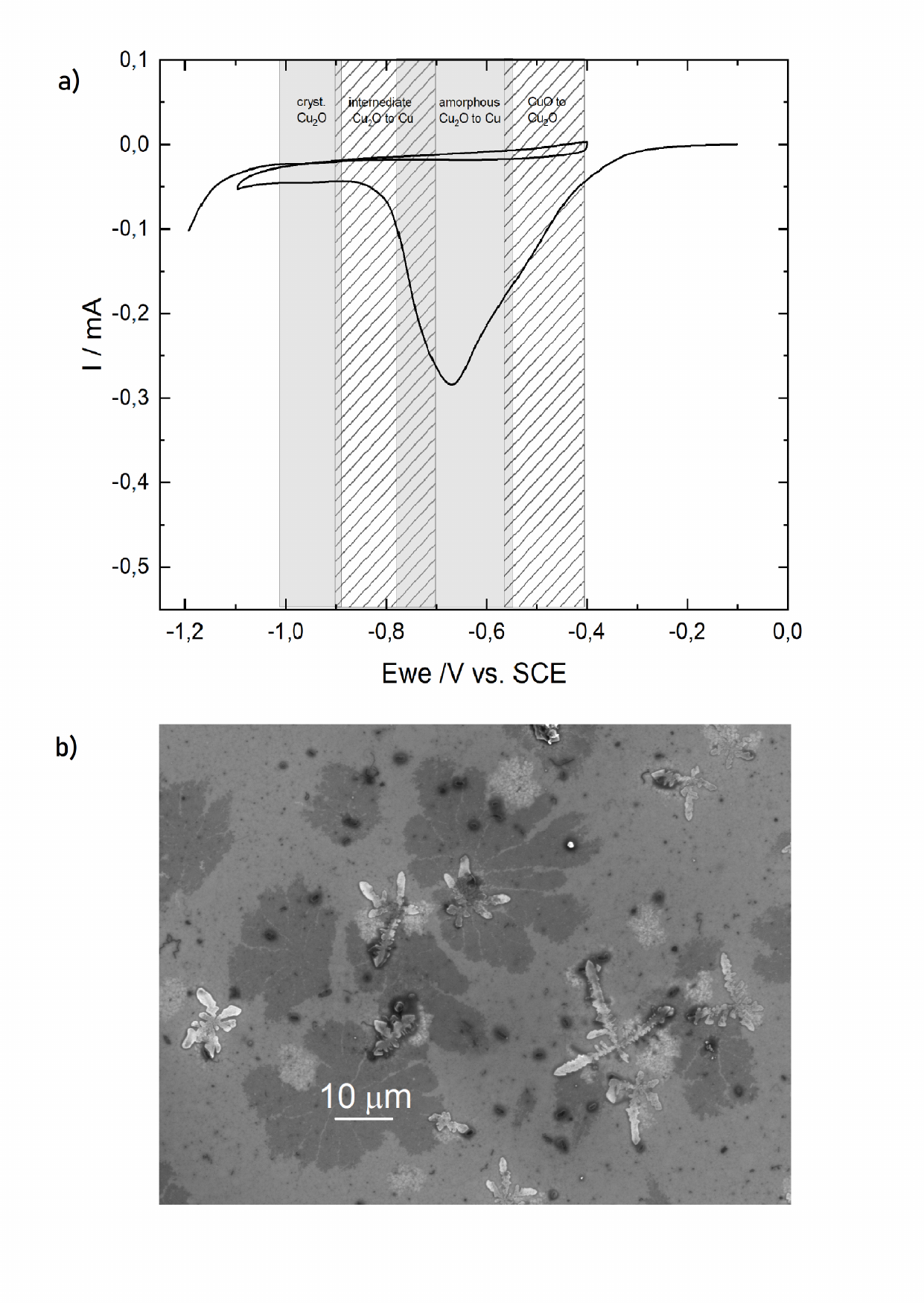}
    \caption{(a) Voltammogram of a copper sample treated with electrolyte, namely 0.05 mM KOH (b) SEM image of a copper sample treated with the same electrolyte.}
    \label{fig:KOH} 
\end{figure}

Lastly, we address the surface modification through plasma-activated electrolytes. Fig. \ref{fig:KOH} shows a CV curve of a copper sample treated with 0.05 mM KOH. The peak for reduction is centered at -0.6 $E_{we}$, indicating an amorphous oxide in an intermediate state similar to the CV curve of an untreated sample (see Fig. \ref{fig:pureCu}). 
However, in the SEM image, some crystals can be seen. They are dendritic rather than cubic. One may argue that these dendrites consist of potassium hydroxide, which cannot be seen in the CV curves since their reduction potentials are outside the considered range.

\begin{figure}[h]
    \centering
    \includegraphics[width=0.6\textwidth]{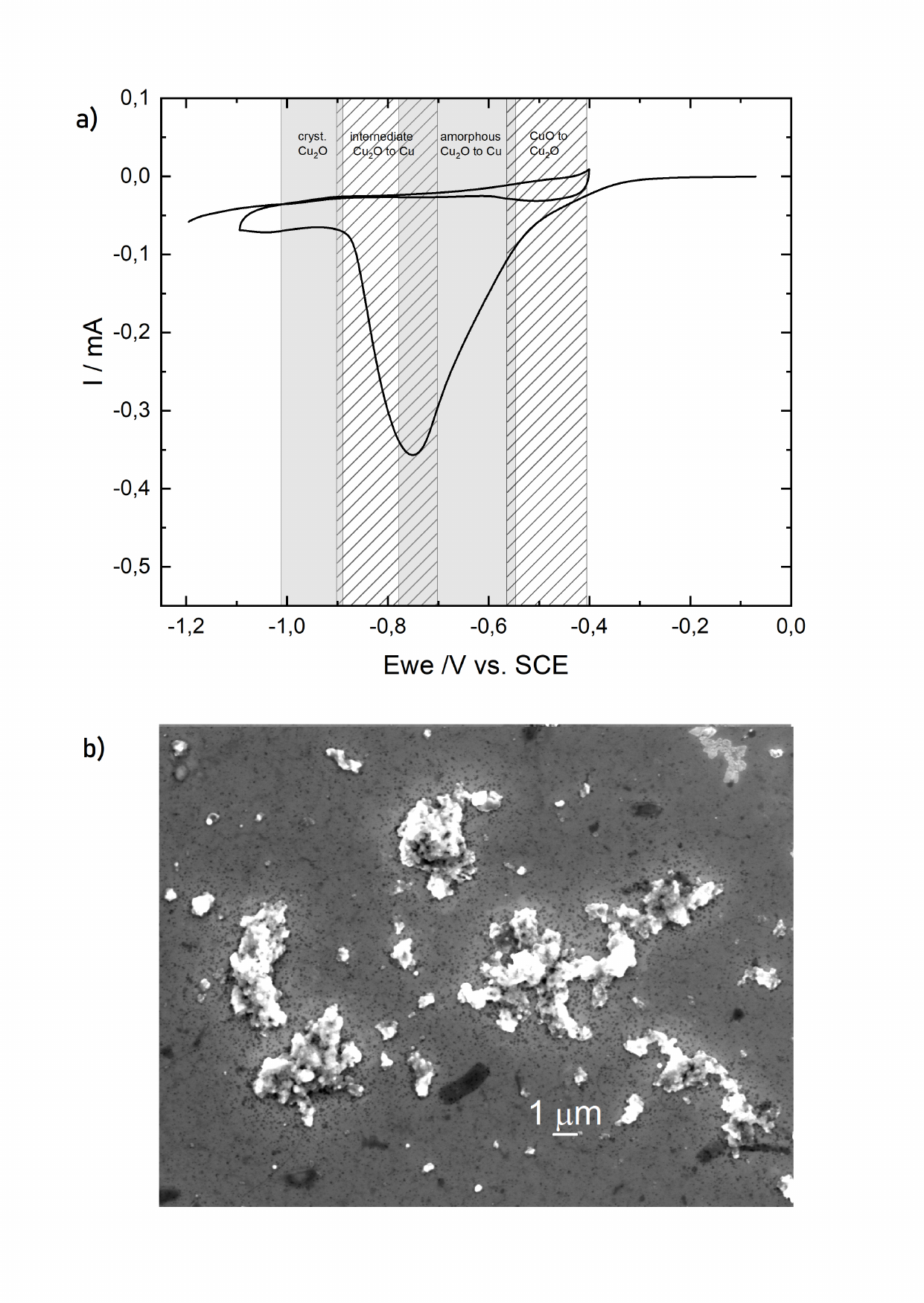}
    \caption{(a) Voltammogram of a copper sample treated with plasma-activated electrolyte, namely 0.05 mM KOH (b) SEM image of a copper sample treated with the same solution.}
    \label{fig:PAE}    
\end{figure}

Fig. \ref{fig:PAE} shows the CV curve for the sample treated with plasma-activated KOH with a concentration of 0.05 mM. The nanosecond-pulsed setup was used in this case. The reduction peak at -0.8 $E_{we}$ indicates the presence of Cu$_2$O in the intermediate state. 
Again, the SEM images show some crystal formation. They have an irregular shape and are smaller than the dendritic crystals formed by the untreated electrolyte. They may also be potassium oxide rather than copper oxide.

In summary, the liquid treated with either the ns- or $\mu$s-pulsed plasma induces the formation of Cu$_2$O in an intermediate state with a small concentration of crystalline Cu$_2$O. These contributions are not stable and are removed in the second electrochemical cycle. This is linked to the very thin oxide layers.

\section{Conclusions}
In this work we have shown, that it is possible to generate uniform copper oxide crystals from both nanosecond- and microsecond pulsed plasmas in liquids using cyclic voltammetry and SEM imaging. The generation of these crystals is initiated by a number of reactive species, hydrogen peroxide being a crucial reactant. The microsecond-pulsed discharge yields a lower H$_2$O$_2$ production, presumably due to its higher dissipated energies. However, the production is still high enough to elicit crystal formation and since the microsecond setup is more versatile and more widely used by other groups it can be a viable tool for the purpose of Cu$_x$O crystal generation. 

However, the microsecond-pulsed plasma causes an increase in temperature and conductivity in the treated liquid as opposed to the nanosecond-pulsed plasma, where these properties show slight or no changes. This causes the $\mu$s-plasma to transition from a standard streamer-like mode into a spark mode, where the temperature of the liquid rapidly increases and much of the previously produced H$_2$O$_2$ is destroyed. The transition point seems to depend on many factors such as conductivity and electrode distance. To optimize H$_2$O$_2$ production and therefore the reactivity of the liquid it is necessary to approach this critical point as close as possibly by regulating factors like treatment time and frequency while never surpassing the transition point. Further investigation is needed to better predict and avoid the transition into spark mode and optimize the reactivity.

Lastly the ns-pulsed plasma was also used to treat a low concentration KOH electrolyte to approach an electrochemical use case. Since it is uncommon to perform electrochemistry in distilled water, it is relevant to know if the Cu$_x$O crystal generation is also possible in such an environment. Admittedly, the ignition of the nanosecond plasma is limited by the conductivity of the electrolyte, therefore the KOH is highly diluted. We have found, that the plasma does change the reactivity of the electrolyte and has an influence on the crystal formation. Again further measurements need to be performed using the microsecond setup, where it is possible to ignite the plasma at higher KOH concentrations.

\newpage
\clearpage

\ack

The DFG has funded this work within the framework of project CRC 1316. Thanks are due to Gabriel Boitel-Aullen and Sascha Chur for the assistance in running the cyclic voltammetry setup and Steffen Schüttler and Sabrina Klopsch for their help regarding the photometric spectroscopy of the PAL. There are no conflicts of interest and no ethical concerns.


\newpage
\clearpage

\printbibliography

\end{document}